\documentclass{article}
\usepackage{ijcai21}

\usepackage{times}

\usepackage{soul}
\usepackage{url}
\usepackage[hidelinks]{hyperref}
\usepackage[utf8]{inputenc}
\usepackage[small]{caption}
\usepackage{graphicx}
\usepackage{amsmath}
\usepackage{booktabs}
\usepackage[table,xcdraw]{xcolor}
\usepackage{multirow}
\title{Major Depressive Disorder Recognition and Cognitive Analysis Based on Multi-layer Brain Functional Connectivity Networks}

\author{
Xiaofang Sun$^1$\and
Xiangwei Zheng$^1$\footnote{Corresponding Author}\and
Yonghui Xu$^{2}$\footnote{Corresponding Author}\and
Lizhen Cui$^{2}$\And
Bin Hu$^1$
\affiliations
$^1$School of Information Science and Engineering, Shandong Normal University.\\
$^2$Joint SDU-NTU Centre for Artificial Intelligence Research (C-FAIR), Shandong University.\\
\emails
\{xiaofangsun2019,~xwzhengcn\}@163.com,
xu.yonghui@hotmail.com,
clz@sdu.edu.cn,
binhu@sdnu.edu.cn.
}

\begin{document}

\maketitle

\begin{abstract}
On the increase of major depressive disorders (MDD), many researchers paid attention to their recognition and treatment. Existing MDD recognition algorithms always use a single time-frequency domain method, but the single time-frequency domain method is too simple and is not conducive to simulating the complex link relationship between brain functions. To solve this problem, this paper proposes a recognition method based on multi-layer brain functional connectivity networks (MBFCN) for major depressive disorder and conducts cognitive analysis. Cognitive analysis based on the proposed MBFCN finds that the Alpha-Beta1 frequency band is the key sub-band for recognizing MDD. The connections between the right prefrontal lobe and the temporal lobe of the extremely depressed disorders (EDD) are deficient in the brain functional connectivity networks (BFCN) based on phase lag index (PLI). Furthermore, potential biomarkers by the significance analysis of depression features and PHQ-9 can be found.
\end{abstract}

\section{Introduction}
Major depressive disorder (MDD) is a universal affective disorder that may cause various mental, physical and cognitive symptoms, and severely hinder the development of the individual's physical and mental health and major social functions \cite{choi2021serum}. With the high incidence and low recognition rate of depression, study on efficient, objective and accurate evaluation methods or biomarkers is an important way to improve public health. The important physiological signal to detect MDD is Electroencephalography (EEG). EEG is the cerebral cortex signal captured by the electrodes of the brain-computer interface device after emotional stimulation. These signals are the product of the gaze of brain neurons after synapse activation. The evaluation of graph theory and the brain functional connectivity network constructed using the EEG electrodes has become a powerful tool to help study MDD recognition.

Recently,~\cite{liu2020functional} used the PLI method to calculate the connectivity matrix and the graph theory-based method to measure the topological structure of the BFCN between different frequency bands in patients with depression. Ding et al. used a multimodal machine learning method, including EEG, eye tracking, and dynamic skin response data as input to classify depression patients and healthy controls \cite{ding2019classifying}.  However, most of these methods use single-layer networks to simulate the associations between brain functions, which is not conducive to in-depth analysis of the complex relationships between brain functions and exploring the cognitive and neural mechanisms that lead to MDD.

This paper mainly explores the cognitive and neural mechanism of MDD and searches for possible biomarkers that can help diagnosis. In order to study the difference between MDD and normal control (NC) in the BFCN, the preprocessed EEG signals are decomposed into Theta, Alpha, Beta1, Beta2, Alpha-Beta1 and All bands frequency bands using wavelet packet transform (WPT), and PLI and phase locked value (PLV) are employed to construct the BFCN, respectively. In order to make up for the limitation and instability of the single BFCN to capture the phase, features extracted from two BFCN are fused at feature level, then the fused features and single-layer features are fed into support vector machine (SVM) for MDD recognition, respectively. After the analysis of the BFCN of NC and MDD, the cognition and understanding of the neural mechanism of depression can be obtained.
The main contributions are as follows:

\begin{figure*}[htb]
	\centering
	\includegraphics[width=0.90\linewidth]{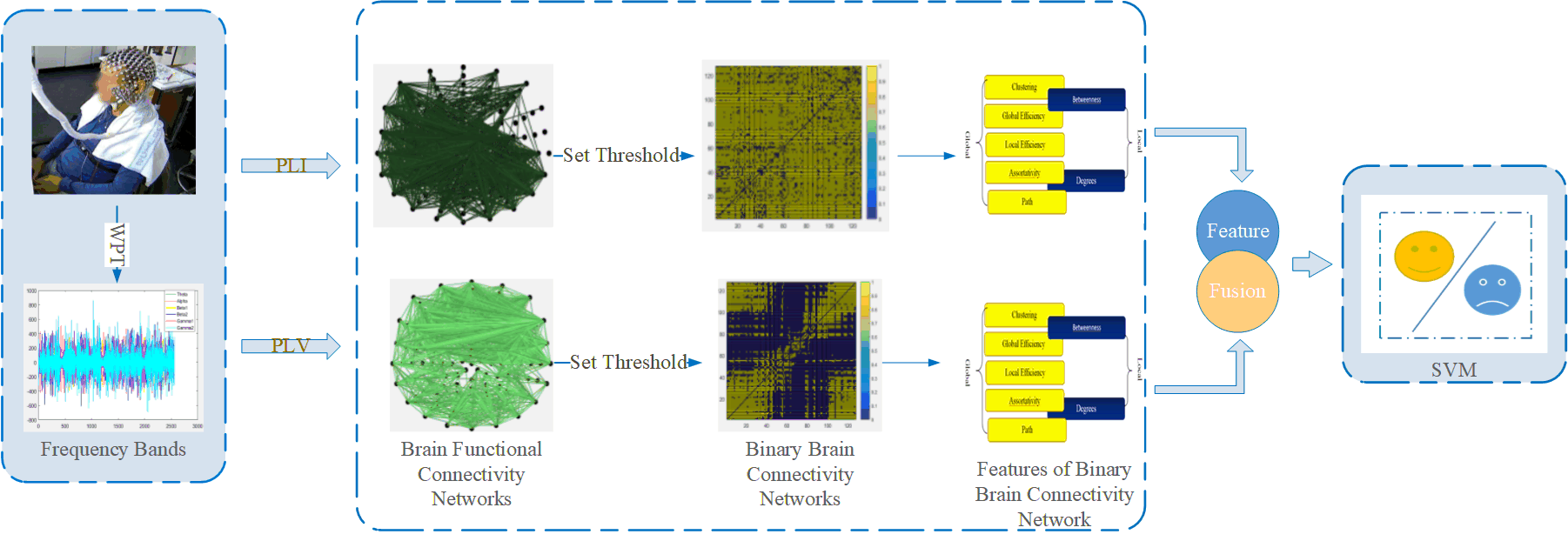}%
	\caption{The framework of MDD recognition based on multi-layer BFCN.} \label{Fig1}
	\vspace*{-5mm}
\end{figure*}
\begin{itemize}
    \item A recognition method based on multi-layer BFCN for MDD is proposed, which fuses the features of the BFCN based on PLI and PLV at the feature level, and improve the stability of the single-layer brain functional connectivity network in MDD recognition. Based on the recognition analysis of the results, we found that Alpha-Beta1frequency band is the key sub-band for recognizing MDD.
    \item Cognitive analysis indicates that the connections between the right prefrontal lobe and the temporal lobe of the extremely depressed disorders (EDD) are deficiency in the BFCN based on PLI. And, potential biomarkers by the significance analysis of depression features and PHQ-9 can be found.
\end{itemize}

\section{The Proposed Method}
To recognize the MDD, we first preprocess EEG signals of each subject, and divide them into Theta, Alpha, Beta1, Beta2 and Alpha-Beta1. Then we apply PLI and PLV to construct the BFCN. After that, we adopt threshold setting method of tensor to transform the brain functional connectivity network into a binary brain network. Then, we extract the five global features (i.e., clustering coefficient, average shortest path length, assortivity coefficient, global efficiency and local efficiency) and two local features (betweenness centrality and node degree) of the binary brain network. With the features, we develop the feature-level fusion algorithm to merge the features of the brain functional connectivity network based on PLI and PLV. At last, we apply base learning (i.e., SVM) to recognize the MDD of the single and the fused multi-layer BFCN.

Considering the length of the original EEG signals and the stress and fatigue of the participants in the early stage of the experiments, EEG signals of the original MDD and NC are normalized and the one-minute EEG signals in the middle of the experiment are extracted as the experimental signals which are subsequently decomposed into 6 frequency bands using WPT db20 wavelet coefficients, namely Theta (4-7Hz), Alpha (8-12Hz), Beta1 (13-19Hz), Beta2 (20-30Hz), Alpha-Beta1 (8-19Hz), All bands (4-60Hz).

\subsection{Multi-layer BFCN Based on PLI and PLV}
Each EEG electrode of each NC and MDD is regarded as a node in the brain network graph, while PLI and PLV are used to construct the edge of the node connections in the BFCN.
PLI is a method to measure the asymmetry of the phase difference distribution between two signals, and is more sensitive to the phase synchronization level \cite{stam2007phase}.
Assume that the instantaneous phases of the signals $x(t)$ and $ y(t)$ of any two channels are $\phi_{x}(t)$ and $\phi_{y}(t)$, respectively.
\begin{equation}
\begin{aligned}
\Delta \varphi(\mathrm{t})=\left|\varphi_{\mathrm{x}}(\mathrm{t})-\varphi_{\mathrm{y}}(\mathrm{t})\right|\\
\end{aligned}
\end{equation}

\begin{equation}
\begin{aligned}
\mathrm{PLI}=\left|<\operatorname{sign}\left[\Delta \varphi\left(\mathrm{t}\right)\right]>\right|
\end{aligned}
\end{equation}

In order to calculate the phase changes of each electrode of the brain, PLV is calculated to capture the nonlinear phase synchronization.
The PLV value is calculated as the connection between any two electrode signals in the time series t, which is the edge of the brain functional connectivity network:
\begin{equation}
\begin{aligned}
\mathrm{PLV}=\left|\frac{1}{\mathrm{N}}\right| \sum_{\mathrm{j}=0}^{\mathrm{N}-1} \mathrm{e}^{\mathrm{i} \Delta \varphi(\mathrm{t})}\\
\end{aligned}
\end{equation}
where N is the number of the electrodes.

\subsection{Binary Brain Network}
In order to reduce the impact of weak connections on the features of the BFCN, the threshold setting of tensor is used to expand a BFCN into 10 sub-binary brain networks.
Binary brain network is constructed as follows.

Step1: Set the initialization threshold;

Step2: Each brain functional connectivity network based on PLI and PLV of six frequency bands of all subjects is traversed to obtain the connections, respectively;

Step3: Set the number of cycles of each brain functional connection network k;

Step4: Judge whether the weight of the connection edge in the BFCN is greater than or equal to the threshold. If it is greater than or equal to the threshold, set the weight of the connection between the two nodes as 1, that is, there is a connection between the two nodes. Otherwise, set to 0, that is, there is no connection between the two nodes, and a sub-binary brain network is obtained.

Step5: Repeat step4 to obtain all sub-binary brain networks under a threshold based on PLI and PLV.

Step6: Reset the threshold and increase to generate a new threshold, repeat steps 2 to 5, and continue to loop until generating all the sub-binary brain networks;

\subsection{MDD Recognition}
After constructing a binary brain network, from the global and the local perspective of the brain network, five global features and two local features are adopted. The five global features include clustering coefficient, average shortest path length, assortativity coefficients, global efficiency, and local efficiency. Two local features include betweenness centrality and node degree.

\begin{figure}[ht!]
	\centering
	\includegraphics[width=\linewidth]{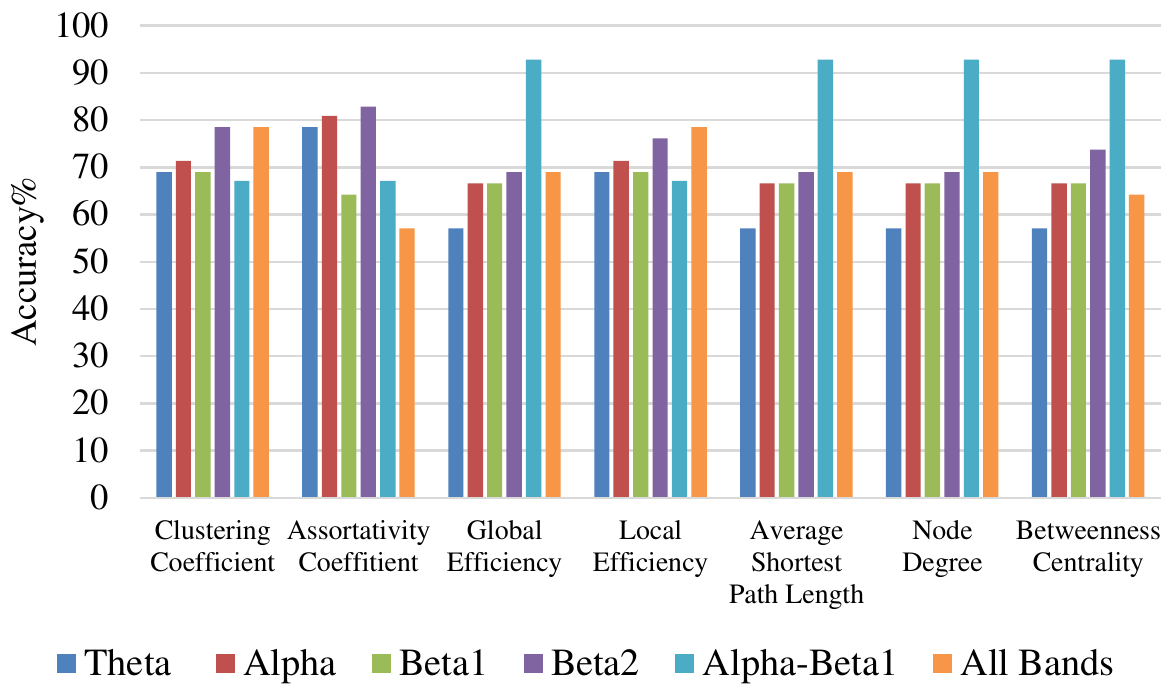}%
	\vspace{-2mm}
	\caption{ MDD recognition result based on PLI BFCN.} \label{Fig2}
	\vspace{-4mm}
\end{figure}
PLI calculates the degree of phase synchronization between any two electrodes at instantaneous t. The main advantage of PLI is that it is not sensitive to volume conduction effect, but it seems to be sensitive to noise. PLV measures the phase difference between the two channels, but PLV is sensitive to volume conduction. Feature level fusion of the BFCN constructed by PLI and PLV can better capture the phase between any two electrodes and reflect the mechanism of the BFCN. The advantage of PLI insensitivity to volume conduction effect can be used to reduce the deficiency of PLV sensitivity. The advantage of PLV insensitivity to noise can be used to compensate for the deficiency of PLI sensitivity.

\begin{equation}
\begin{aligned}
 F = \left\{ {     \mathrm{PLI}\{\mathrm{C}, \mathrm{L,P}, \mathrm{E}_{\mathrm{Global}}, \mathrm{E}_{\text {local }}, \mathrm{B}, \mathrm{K}_{\mathrm{N}}\}      } \right. \\
 \left. {        \mathrm{PLV} \{\mathrm{C}, \mathrm{L,P}, \mathrm{E}_{\text {Global }}, \mathrm{E}_{\text {local }}, \mathrm{B}, \mathrm{K}_{\mathrm{N}} \}                 } \right \} \\
 \end{aligned}
\end{equation}

In the field of MDD recognition, SVM show good prediction and recognition performance. The kernel function of SVM adopts the radial basis kernel function (RBF), and the learning method adopts the sequential minimum optimization (SMO) method.

\section{Experiments and Analysis}
This study uses the public dataset MODMA~\cite{sun2019graph} to construct the single-brain and the multi-BFCN to recognize MDD. A total of patients diagnosed with depression (female/male=11/13, 30.88$\pm$10.37 years old) 24 MDD were used as experimental subjects, and 29 NC (female/male=9/20, 31.45$\pm$9.15 years old) were used as control groups. There was no significant difference in age (t=0.214, p=0.832) or gender (x2=1.224, p=0.269) between groups.

\subsection{Experiments on MDD Recognition}
Figure 2 shows the MDD recognition result with the BFCN constructed based on PLI. The hightest accuracy in Figure 2 is 92.86\%. As shown in Figure 3, the highest recognition accuracy of the average features of the BFCN constructed based on PLV is 88.10\%. It can be concluded that the BFCN constructed by PLI has a strong ability to recognize MDD and NC .
As we can see in Figure 4, feature level fusion of PLI and PLV-based multi-layer BFCN improves the stability of the MDD recognition, and shows good recognition results in six frequency bands, and the highest MDD recognition result in the Alpha-Beta1 frequency band is 92.86\%.

\begin{figure}[ht!]
	\centering
	\includegraphics[width=\linewidth]{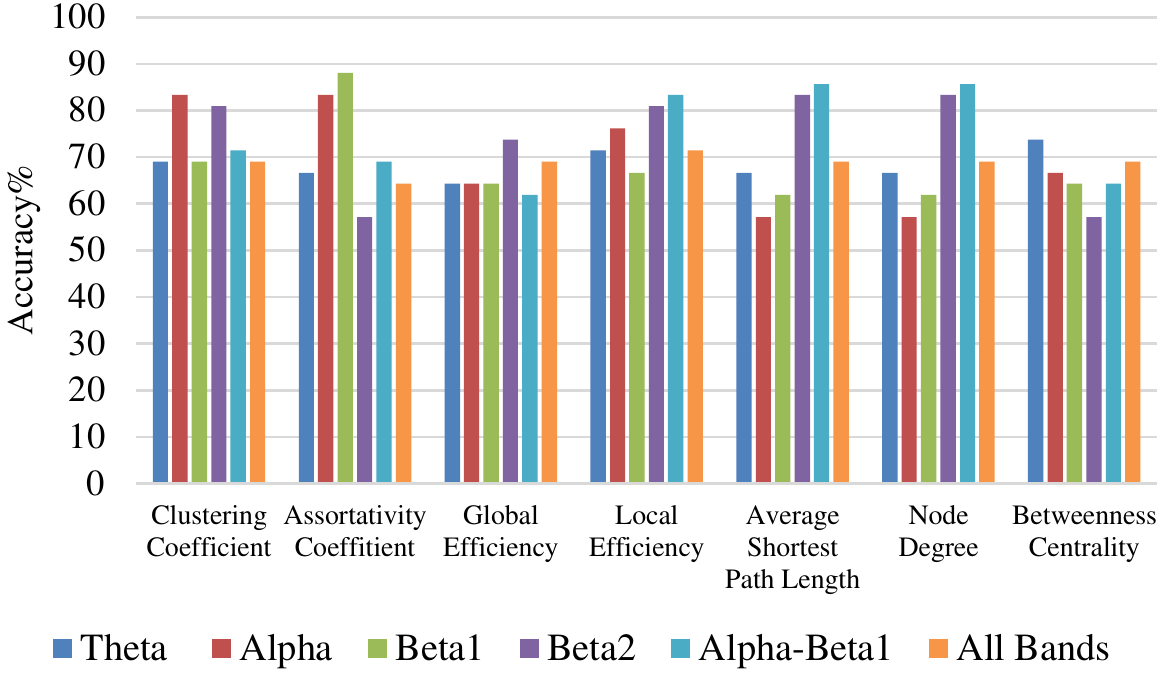}%
	\vspace{-2mm}
	\caption{ MDD recognition result based on PLV BFCN.} \label{Fig3}
\end{figure}
\begin{figure}[ht!]
	\centering
	\includegraphics[width=\linewidth]{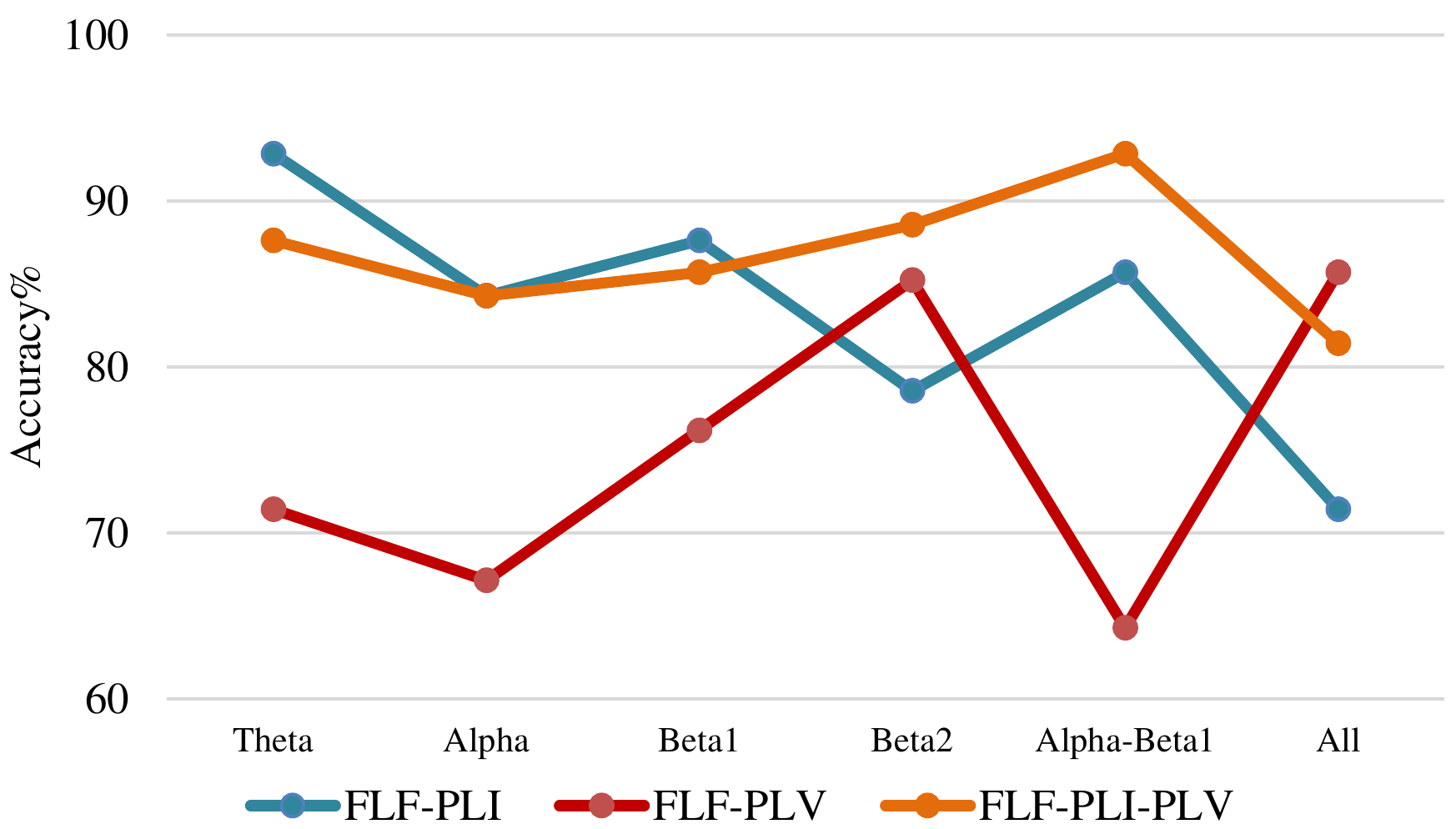}%
	\vspace{-2mm}
	\caption{Comparison result between single-layer and multi-layer BFCN on MODMA.} \label{Fig4}
	\vspace{-4mm}
\end{figure}

\subsection{Cognitive Analysis on MDD}
In Figure 5, the single-layer BFCN based on PLI and PLV achieved the highest average recognition effect compared with other frequency bands in the Alpha-Beta1 frequency band. It also performs best in the MDD recognition of multi-layer BFCN. Then, it can be determined that the Alpha-Beta1 frequency band is the key sub-band.

To explore the brain functional connections between MDD and NC in each brain area of the BFCN in the resting state to further recognize the neural mechanism of MDD, the PHQ-9 score is further divided, PHQ-9$<$5 is NC, 5$\le$PHQ-9$<$18 is MDD, PHQ-9$>$=18 is extreme depressive disorder (EDD). As shown in Figure 6, in the brain functional connectivity network based on PLI, all areas of the NC brain functional connectivity network are connected in a resting state, and the NC brain functional connectivity network is normal without damage. There is a connection deficiency in the right frontal and temporal lobe regions of the brain functional connectivity network of MDD, which means that the brain functional connectivity network of MDD has cognitive deficiency in the resting state. As the degree of depression increases, the connection deficiency of the brain functional connectivity network of EDD increases. There are connection deficiencies in the left frontal lobe, right frontal lobe, and right temporal lobe. In general, the increased degree of depression has an increased cognitive impact on MDD, and MDD has cognitive deficits in the frontal and temporal lobes of the right brain.

\begin{figure}[htb]
	\centering
	\includegraphics[width=\linewidth]{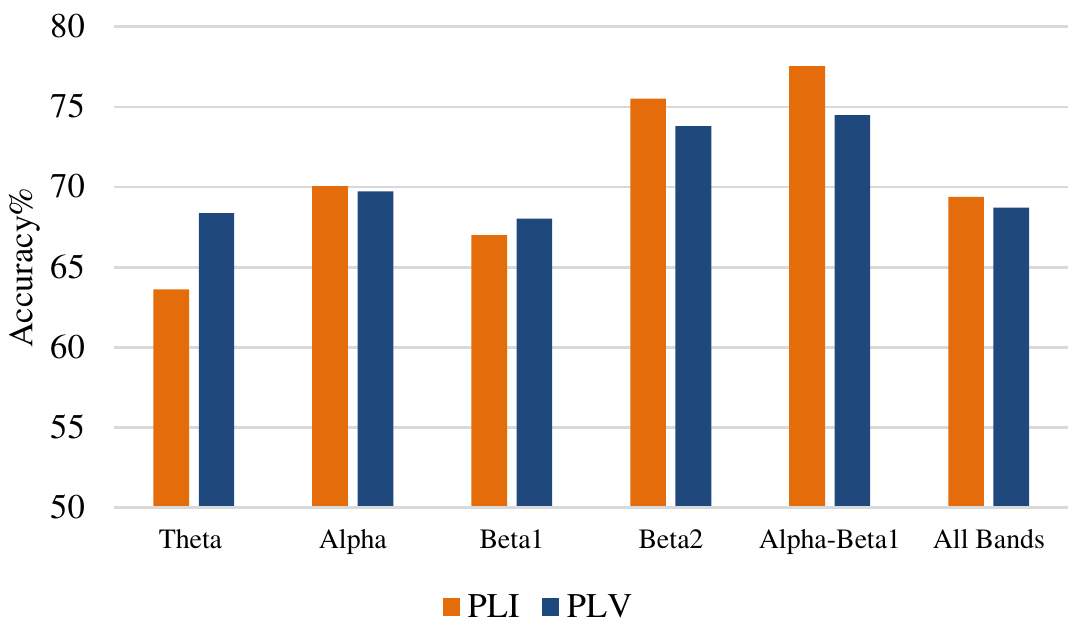}%
	\vspace{-3mm}
	\caption{ MDD recognition result of single-layer BFCN with respect to the six frequency bands.} \label{Fig5}
	\vspace{-3mm}
\end{figure}
\begin{table}
	\centering
	\begin{tabular}{lll}
		\hline
		Reference  &  Method  &  Accuracy\\
		\hline
		\cite{sun2020study}      &  ReliefF& 82.31\%      \\
		\cite{sun2019graph}      & Coh/ICoh/Corr/PLI/PLV & 87.50\%       \\
		Proposed Method      & PLI+PLV &  92.86\%      \\
		\hline
	\end{tabular}
	\vspace{-1mm}
	\caption{Comparison results on MODMA.}
	\label{tab:plain}
	\vspace{-4mm}
\end{table}

In the BFCN based on PLV, there is no obvious difference between NC, MDD and EDD.
To further explore the neural mechanism of MDD, the analysis is carried out from the perspective of features of six frequency bands, and the significance of the seven features of the six frequency bands and the PHQ-9 value is analyzed.
According to the P value of one-way ANOVA of seven features in six frequency bands based on PLI and PHQ-9, it can be seen that the features of the six frequency bands and PHQ-9 score of the BFCN based on PLI all conform to P$<$0.005, showing significant differences. Then, all features of PLI-based BFCN could be biomarkers. According to the P value of the one-way ANOVA of seven features in six frequency bands based on PLV and PHQ-9, it can be seen that there is statistical significance except for the feature of degree in the ALL band and the analysis result of PHQ-9, which is P=0.199 (p$>$0.005). The features and PHQ-9 scores of other PLV-based BFCN were in line with P$\le$ 0.005, showing significant differences. Therefore, PLV-based BFCN and PHQ-9 features that are significantly different may be biomarkers.

\section{Conclusion}
This research proposes a MDD recognition method based on multi-layer BFCN. The BFCN features based on PLI and PLV are fused using feature level fusion algorithms, and finally the features of single-layer and multi-layer BFCN are input into the SVM model to recognition MDD and NC. Experimental results show that the best recognition accuracy of the proposed multi-layer brain functional connectivity network fusion method is 92.86\%. These results prove the complementary features of the multi-layer BFCN. At the same time, the connections between the right frontal lobe and the temporal lobe of MDD are deficiency, and it provides potential biomarkers for MDD recognition.

\begin{figure}[ht]
	\centering
	\includegraphics[width=0.90\linewidth]{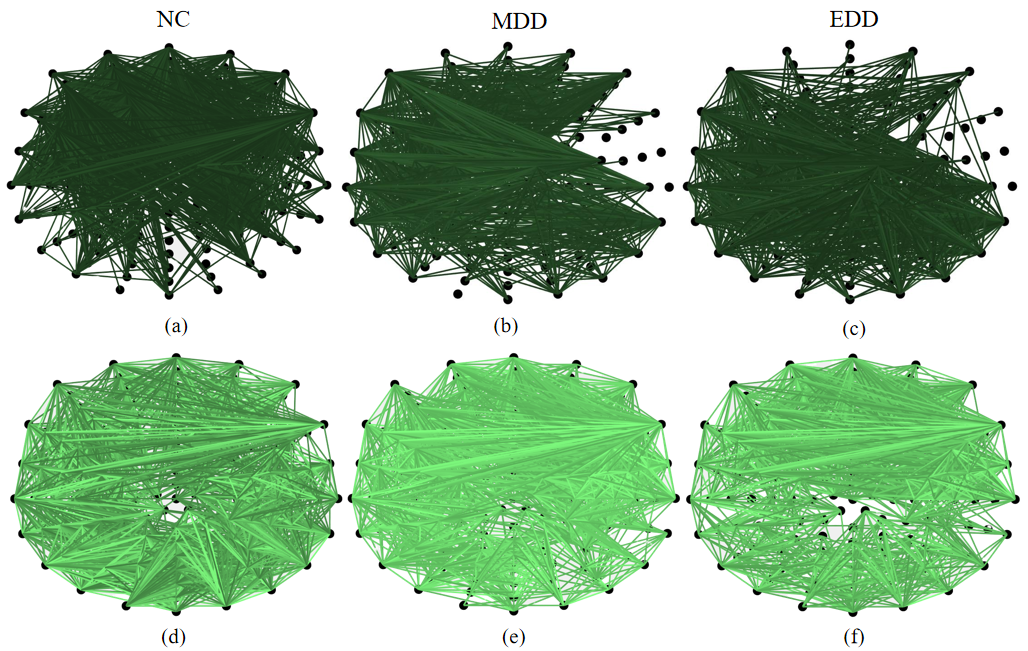}%
	\caption{ Comparison of the BFCN of NC, MDD and EDD based on PLI and PLV in key sub-bands Alpha-Beta1.
		Figures (a), (b), (c) are brain functional connection networks based on PLI, and Figures (d), (e), (f) are brain functional connection networks based on PLV.} \label{Fig6}
	\vspace{-4mm}
\end{figure}
\section{Acknowledgment}
We are grateful for the support of the National Natural Science Foundation of China (91846205) and the Natural Science Foundation of Shandong Province (No.ZR2020LZH008, ZR2020QF112, ZR2019MF071) and the Fundamental Research Funds of Shandong University.

\bibliographystyle{named}
\bibliography{ijcai21}

\end{document}